\def \be {\begin{equation}}
\def \ee {\end{equation}}
\def \bea {\begin{eqnarray}}
\def \eea {\end{eqnarray}}
\def \nn {\nonumber}

\def \la {\langle}
\def \ra {\rangle}

\def \del {\partial}
\def \dels {\partial\kern-.5em / \kern.5em}
\def \As {{A\kern-.5em / \kern.5em}}
\def \Ds {D\kern-.7em / \kern.5em}

\def \a {\alpha}
\def \b {\beta}

\def \G {\Gamma}
\def \d {\delta}
\def \eps {\epsilon}

\def \lam {\lambda}

\def \s {\sigma}

\def \om {\omega}
\def \Om {\Omega}

\def \th {\theta}

\def \Ah {\hat{A}}
\def \hG {\hat{G}}
\def \Gh {\hat{G}}
\def \Fh {\hat{F}}
\def \Bh {\hat{B}}

\def \Rt {\tilde{R}}

\documentclass[12pt]{article}

\begin{document}
\begin{titlepage}
%\catcode`\@=11
%\catcode`\@=12
%\twocolumn[\hsize\textwidth\columnwidth\hsize\csname%
%@twocolumnfalse\endcsname

%\draft
\begin{center}
%\hfill{NEIP-98-022}
\hfill hep-th/0105191\\
\vskip .5in

\textbf{\large
Noncommutative Differential Calculus \\
for D-brane in Non-Constant B Field Background \\
%and Background Independence for Metric \\
}

\vskip .5in
{\large Pei-Ming Ho, Shun-Pei Miao}
\vskip 15pt

{\small \em Department of Physics, National Taiwan
University, Taipei 106, Taiwan, R.O.C.}

\vskip .2in
\sffamily{
pmho@phys.ntu.edu.tw\\
meowshun@ms28.url.com.tw}

\vspace{60pt}
%\maketitle
\end{center}
\begin{abstract}

In this paper we try to construct noncommutative
Yang-Mills theory for generic Poisson manifolds.
It turns out that the noncommutative differential
calculus defined in an old work is exactly what we need.
Using this calculus, we generalize results about
the Seiberg-Witten map, the Dirac-Born-Infeld action,
the matrix model and the open string quantization
for constant $B$ field to non-constant background with $H=0$.

\end{abstract}
%\pacs{PACS numbers: 11.25.-w, 11.25.Mj, 11.25.Sq}%]
\end{titlepage}
%\begin{narrowtext}
\setcounter{footnote}{0}

\section{Introduction} %\hspace{5pt}

For a D-brane in a constant B field background,
its low energy effective theory lives
on a noncommutative space for which
the commutator
\be \label{xxth}
[x^a, x^b]_{\ast}=i\th^{ab}
\ee
is a constant \cite{NC}.
In the zero slope limit of Seiberg and Witten \cite{SW}
\be \label{zsl}
\a'\sim\eps^{1/2}, \quad
g_{ab}\sim\eps,
\ee
where $\eps\rightarrow 0$,
and $B_{ab}$ is fixed,
we have $\th=B^{-1}$.
If the B field background is not constant,
but has a vanishing field strength $(H=dB)$
for longitudinal directions on the D-brane,
$\th$ satisfies the Jacobi identity and so it
defines a Poisson structure on the D-brane \cite{HY,CS,Sch}.
One can then use Kontsevich's formula \cite{Kon}
to define the $\ast$-product as a way
to quantize the Poisson structure,
and to define the noncommutative D-brane worldvolume.
If $H\neq 0$, the Kontsevich's formula can still
be used to construct integrations to reproduce
open string correlation functions \cite{CS},
however the resulting algebra is non-associative.
An associative algebra was found by
quantizing the open string \cite{HY},
and it can also be used to reproduce
the open string correlation functions \cite{Ho2}

In this paper we try to construct
the noncommutative Yang-Mills theory
for non-constant $B$ with $H=0$.
For simplicity, we ignore terms of order
$\th^2$ or higher, that is,
we work in the Poisson limit.
Interestingly, it turns out that an old work \cite{CH}
on the formulation of differential calculus
on noncommutative spaces is precisely
what is needed for our purpose, i.e.,
to define the Seiberg-Witten map,
and to find a noncommutative Yang-Mills (NCYM) theory
that approximates the classical Dirac-Born-Infeld (DBI)
action with the $B$ field background.

The first problem one meets when trying to
define a field theory on noncommutative space
is to define derivatives.
When $\th$ is constant, we usually assume that
\be
[\del_a, x^b]_{\ast}=\delta_a^b.
\ee
However, when $\th$ is not constant,
this relation is inconsistent with (\ref{xxth}),
as can be easily seen by checking the Jacobi identity
for $(\del_a, x^b, x^c)$.

Another way to state the same problem is to notice
that although we can define differential forms $(dx^a)$
satisfying
\be \label{1}
[x^a, dx^b]_{\ast}=0
\ee
when $\th$ is constant,
after a general coordinate transformation
this relation no longer holds,
even for those transformations
which keep $\th$ constant.
Thus even for a constant $\th$
(which you can always achieve by coordinate transformations),
it is unclear how to determine $[x^a, (dx^b)]_{\ast}$
or $[\del_a, x^b]_{\ast}$ for a generic curved space.

One might suspect that, in the $\ast$-product
representation of the algebra,
the classical derivatives $\del_a$ can be used
as the natural representative of
the derivatives on noncommutative space.
Although this is an algebraically consistent choice,
we find that this definition is not enough.
If the gauge transformation for a gauge potential is defined as
\be
\hat{\d}\hat{A}_a=\del_a\hat{\lam}-i[\hat{A}_a,\hat{\lam}]_{\ast},
\ee
the field strength defined by
\be
\hat{F}_{ab}=
\del_a\hat{A}_b-\del_b\hat{A}_a-i[\hat{A}_a,\hat{A}_b]_{\ast}
\ee
will not be covariant.
The reason is basically that the Leibniz rule fails
\be
\del_a(f\ast g)\neq(\del_a f)\ast g+f\ast(\del_a g).
\ee

In the context of matrix model,
the generalized notion of covariant derivative,
the covariant coordinates $X=x+A$,
was introduced \cite{Li,SW}.
% new
In particular the problems mentioned above were
resolved in \cite{Schupp,JSW} by using only
the covariant coordinates without refering to the derivatives.
Thus although some results of this paper overlap
with \cite{JSW}, here
% new
we are interested in a more conventional description
of ncommutative gauge field theory,
because this subject is of interest by itself
apart from its application in string theory.

We review our mathematical formulation
of noncommutative differentail calculus
at the Poisson level in sec.\ref{PDC}.
In sec.\ref{NGT} we define the NCYM action,
find the Seiberg-Witten map,
and check that after suitable modification,
the NCYM action agrees with the DBI action.
In sec.\ref{Remarks}
we provide an interpretation of the noncommutative calculus
from the viewpoint of open string quantization.
We also comment on the issue of background independence
and general coordinate transformations.

\section{Poisson Differential Calculus} \label{PDC}

\subsection{Definition} \label{definition}

In this section and the next we briefly summarize
part of the results of \cite{CH}.

To discuss differential calculus on a noncommutative space
at the lowest order approximation,
we define the Poisson structure on differential forms as follows.
(Its dual description in terms of derivatives
will be given in sec.\ref{Dual}.)
The differential calculus on a noncommutative space
is generated by the coordinates $x$ and one-forms $dx$.
We generalize the Poisson bracket on $x$
\be \label{xx}
(x^a, x^b)=\th^{ab}(x)
\ee
to any two differential forms such that
\be \label{sym}
(\om_1, \om_2)=(-1)^{p_1 p_2+1}(\om_2, \om_1),
\ee
where $p_i$ is $0$ or $1$ if $\om_i$ is an even or odd form.
After quantization, the Poisson bracket
is an approximation to the commutator or anti-commutator,
depending on whether the two differential forms
commute or anti-commute classically.
For instance,
\bea
[ x^a, x^b ]_{\ast} &\simeq& i(x^a, x^b)+\cdots, \\
{[ x^a, dx^b ]}_{\ast} &\simeq& i(x^a, dx^b)+\cdots, \\
\{dx^a, dx^b\}_{\ast} &\simeq& i(dx^a, dx^b)+\cdots.
\eea
The $\ast$-product to the lowest order in $\th$ is
\be
\om_1\ast\om_2 = \om_1\om_2+\frac{i}{2}(\om_1,\om_2)+\cdots.
\ee

The associativity of the quantized algebra implies that
the Poisson brackets should satisfy the graded Jacobi identities
\be
\sum_{(i,j,k)}(-1)^{p_i p_k}(\om_i,(\om_j,\om_k))=0,
\ee
where we sum over cyclic permutations of $(i,j,k)$.
The Leibniz rule for the commutators and anti-commutators
of the quantized algebra also implies that
\be
(\om_1,\om_2\om_3)=
(\om_1,\om_2)\om_3+(-1)^{p_1 p_2}\om_2(\om_1,\om_3).
\ee
In addition, we have the Leibniz rule for the exterior derivative
\be \label{Leib}
d(\om_1,\om_2)=(d\om_1,\om_2)+(-1)^{p_1}(\om_1,d\om_2).
\ee
Finally, we require that if $\om_i$ is a $n_i$-form,
$(\om_1,\om_2)$ is always a $(n_1+n_2)$-form.

Assuming that $\th^{ab}$ is invertible,
we can always write $(x, dx)$ in the form
\be \label{xdx}
(x^a, dx^b)=-\th^{ac}\G^b_{cd}dx^d,
\ee
where $\G^b_{cd}$ are some functions of $x$
which transforms like a connection
under general coordinate transformations.
As $\G$ is in general not symmetric,
one can use
\be
\tilde{\G}^a_b\equiv \G^a_{bc}dx^c
\ee
and
\be
\G^a_b\equiv dx^c\G^a_{cb}
\ee
as the connection one-forms
to define two kinds of covariant derivatives
$\tilde{\nabla}$ and $\nabla$, respectively.
Given $\th$ and $\Gamma$,
all Poisson brackets are determined.

The Leibniz rule for two functions
\be \label{Leib1}
d(x^a,x^b)=(dx^a,x^b)+(x^a,dx^b)
\ee
implies that the Poisson structure is covariantly constant
\be \label{covar-th}
\tilde{\nabla}_a\th^{bc}\equiv
\del_a\th^{bc}+\th^{bd}\G^c_{da}+\G^b_{da}\th^{dc}=0.
\ee
For a function and a one-form, the Leibniz rule
\be \label{Leib2}
d(x^a,dx^b)=(dx^a,dx^b)
\ee
can be viewed as the definition of
Poisson brackets for two one-forms,
except that it has to be consistent with (\ref{sym}), i.e.,
$(dx^a,dx^b)=(dx^b,dx^a)$.
It turns out that this is guaranteed by (\ref{covar-th}),
and explicitly
\bea \label{r}
(dx^a,dx^b)&=&-\frac{1}{2}r^{ab}_{cd}dx^c dx^d \nn\\
&=&-\frac{1}{2}d(\th^{ac}\tilde{\G}^b_c+\th^{bc}\tilde{\G}^a_c).
\eea
The last expression above also implies that
the Leibniz rule for two one-forms
\be \label{Leib3}
d(dx^a,dx^b)=0
\ee
is automatically satisfied.
Thus we see that once (\ref{Leib1}) is satisfied,
all the rest of the Leibniz rules follow.
Intriguingly, we will see in sec.\ref{SW-map} that
the existence of the Seiberg-Witten map
requires that (\ref{Leib1}) holds.
Note that the Leibniz rules are not automatically
satisfied by any associative noncommutative algebra
of $x$ and $dx$ in the Poisson limit.

\subsection{Canonical Coordinates} \label{canonical}

In \cite{CH} it was found that the Jacobi identities
are so strong that locally we can always
make a change of coordinates such that
for the new coordinates $\Phi^a=\Phi^a(x)$
\be \label{phiphi}
(\Phi^a, \Phi^b)\equiv P^{ab}
=\frac{1}{2}\Rt^{ab}_{cd}\Phi^c\Phi^d+T^{ab}_c\Phi^c+\th_0^{ab},
\ee
where $\Rt, T$ and $\th$ are constant tensors
constrained by the Jacobi identity for $P$.
For instance, $\Rt$ has to satisfy
the classical Yang-Baxter relation.
Therefore, we can classify all Poisson differential calculus
by these constant tensors up to linear transformations of $\Phi$.

Letting $x^a=\Phi^a$,
one also finds \cite{CH} that
\be \label{Gamma}
\G^a_{bc}=P^{ad}\del_b P^{-1}_{dc}.
\ee
Since the connection $\G$ is a pure gauge,
it is convenient to define a new basis of one-forms
\be
e_a=P^{-1}_{ab}d\Phi^b,
\ee
for which the algebra is greatly simplified
\bea
(e_a, x^b)&=&0, \\
(e_a, e_b)&=&-\frac{1}{2}\Rt_{ab}^{cd}e_c e_d.
\eea
Note that here $\Rt$, which also appeared in (\ref{phiphi}),
is just the curvature tensor
expressed in the basis of $e_a$
for the connection one-form $\tilde{\Gamma}$,
with its index raised by $P$.
Jacobi identities require that $\Rt$ be constant.
Similarly, $T$ can be understood as
the torsion at the origin $\Phi=0$.

An interesting feature of
the Poisson differential calculus is that
one can always realize the action of the exterior derivative
on any function $f(x)$ through a one-form $\xi$
\be \label{xi}
\xi=-e_a\Phi^a
\ee
as
\be
df=(\xi, f)=-e_a(\Phi^a,f).
\ee
This is a property that will be useful
for the formulation of matrix model.
It extends to all differential forms
\be
d\om=(\xi,\om)
\ee
only if $T=0$.

\subsection{Integration} \label{Connes}

The integration $\la f\ra$ of a function $f$
over the noncommutative space can be defined as
\be
\la f\ra\equiv Tr(f),
\ee
where the trace is taken over a Hilbert space
representation of the algebra of $x$ after quantization.
%(For infinite dimensional representations,
%one may need to use the Dixmier trace. \cite{Con})

For constant $\th$, the trace is
the noncommutative counterpart of integration
\be
|Pf(\th)|Tr \leftrightarrow \int d^D x.
\ee
Under general coordinate transformations $x\rightarrow x'$,
$|Pf(\th)|^{-1}$ transforms by a factor of the Jacobian
$|det(\del x'/\del x)|$, so we can deduce that
\be
Tr \leftrightarrow \int d^D x|Pf(\th^{-1})|
\ee
even when $\th$ is not constant.
Note that in general $|Pf(\th^{-1})|$ is not
the measure $\sqrt{det(G)}$ for a space with metric $G$.

At the Poisson level,
\be
\la\cdot\ra = Tr(\cdot) \simeq \int d^D x|Pf(B)|(\cdot),
\ee
where $B=\th^{-1}$ and $|Pf(B)|=\sqrt{det(B)}$.
Since
\be
\del_a Pf(B)=Pf(B)\th^{-1}_{ac}(\del_b\th^{bc}),
\ee
one can check that
\be
\del_a(Pf(B)\th^{ab})=0.
\ee
Using this relation it is easy to see that
the integration $\la\cdot\ra$ is cyclic in the Poisson limit
\be \label{cyclic}
\la f\ast g\ra = \la g\ast f\ra
\Rightarrow \la(f,g)\ra=0
\ee
for arbitrary integrable functions $f$ and $g$.
This implies that
\be
\la f\ast g\ra\simeq\la fg\ra.
\ee

\section{Noncommutative Gauge Theory} \label{NGT}

\subsection{Noncommutative Gauge Fields}

In this paper, we will mostly work at the Poisson level,
that is, to the first order of $\th$ or $P$.
(The $\Gamma$ defined in (\ref{xdx}) is of the zeroth order.)
Define the gauge transformation
for a noncommutative gauge potential one-form
\be
\hat{A}=dx^a\hat{A}_a(x)
\ee
by
\be
\hat{\d}\hat{A}
=d\hat{\lam}-i[\hat{A}, \hat{\lam}]_{\ast}
\simeq d\hat{\lam}+(\hat{A}, \hat{\lam}).
\ee
It follows from this that
\be
\hat{\d}\hat{A}_a\simeq
\del_a\hat{\lam}+\th^{bc}\nabla_b\Ah_a\del_c\lam,
\ee
where
\be
\nabla_a\Ah_b\equiv \del_a\Ah_b-\Gamma^c_{ab}\Ah_c.
\ee
Define the field strength two-form by
\bea \label{F}
\hat{F}&=&\frac{1}{2}dx^a dx^b\hat{F}_{ab}(x) \nn\\
&=& d\hat{A}-i\hat{A}\ast\hat{A}.
\eea
For $U(1)$ gauge fields,
\be
\hat{F}\simeq d\hat{A}+\frac{1}{2}(\hat{A},\hat{A}).
\ee
More explicitly,
\be
\Fh_{ab}=\del_a\Ah_b-\del_b\Ah_a
+\frac{1}{2}\nabla_c\Ah_a\th^{cd}\nabla_d\Ah_b
-\frac{1}{2}\tilde{\cal R}^{cd}_{ab}\Ah_c\Ah_d,
\ee
where 
\be
\tilde{\cal R}^{cd}_{ab}=\th^{ce}\tilde{\cal R}^d_{eab},
\ee
and
\bea
\tilde{\cal R}^a_b&=&\frac{1}{2}\tilde{\cal R}^a_{bcd}dx^c dx^d \nn\\
&=&d\tilde{\G}^a_b+\tilde{\G}^a_c\tilde{\G}^c_b.
\eea

Under a gauge transformation,
\be
\hat{\d}\hat{F}\simeq(\hat{F}, \hat{\lam}),
\ee
where we used the fact that the Leibniz rule (\ref{Leib}) holds.
In terms of the components
\be \label{dF}
\hat{\d}\hat{F}_{ab}\simeq\th^{cd}\nabla_c\hat{F}_{ab}\del_d\hat{\lam},
\ee
where
\be \label{nablaF}
\nabla_a\hat{F}_{bc}\equiv
\del_a\hat{F}_{bc}-\hat{F}_{bd}\G^d_{ac}-\G^d_{ab}\hat{F}_{dc}.
\ee

We also need the Leibniz rule to show that $\hat{F}$
satisfies the Bianchi identity
\be \label{Bianchi}
d\hat{F}+(\hat{A},\hat{F})\simeq 0.
\ee

The covariant differential one-form is defined as
\be
\hat{D}=d+\hat{A}
=dx^a(\del_a+\hat{A}_a).
\ee
One may also choose to use $e_a$ as the basis of one-forms.
Using (\ref{xi}), we find
\be
\hat{D}\simeq\xi+A=e_a(-\Phi^a+\hat{A}^a)
\ee
when acting on functions of $x$.
So the covariant coordinate $X=x+A$
has a conventional interpretation when $x$
happens to be the canonical coordinate $\Phi$.

\subsection{Seiberg-Witten Map} \label{SW-map}

The Seiberg-Witten map is a map from
a commutative gauge potential $A$ to
a noncommutative gauge potential $\hat{A}$
such that the gauge transformations of $A$
are maped to the gauge transformations of $\Ah$ \cite{SW}.
The Seiberg-Witten map for arbitrary Poisson structure $\th$
was given in \cite{Schupp} for the $A$ field defined
in the covariant coordinates $X=x+A$.
Here we present a map for the $A$ field defined
in the covariant derivative $D=\del+A$.
This map has to match gauge transformations of $A$
to those of $\hat{A}$
\be \label{SW0}
\hat{A}(A)+\hat{\d}_{\hat{\lam}}\hat{A}(A) = \hat{A}(A+\d_{\lam}A).
\ee
Here we only consider $U(1)$ gauge fields,
so $\d_{\lam}A=d\lam$.

Comparing with the Seiberg-Witten map for constant $\th$,
we have a new term on the left hand side of (\ref{SW0})
by differentiating $\th$ in $\hat{\lam}$.
We hope to compensate this change
by choosing the $\Gamma^a_{bc}$ factor in (\ref{xdx})
as well as modifying the Seiberg-Witten map appropriately.
Interestringly, it turns out that we should demand $\G$
to satisfy the Leibniz rule (\ref{Leib1}).
(As we mentioned in the end of sec.\ref{PDC},
this guarantees that all Leibniz rules can be satisfied.)
We modify the Seiberg-Witten map
by adding a new term involving $\G$ as
\bea \label{SWA}
\hat{A}_a&=&A_a-\frac{1}{2}\th^{bc}A_b(\del_c A_a+F_{ca})
+\frac{1}{4}(\th^{bd}\G^c_{da}+\th^{cd}\G^b_{da})A_b A_c+\cdots \nn\\
&=&A_a-\frac{1}{2}\th^{bc}A_b(\nabla_c A_a+F_{ca})+\cdots.
\eea
The transformation parameters $\hat{\lam}$ and $\lam$
are still related by the same map
\be \label{SWlam}
\hat{\lam}=\lam+\frac{1}{2}\th^{ab}\del_a\lam A_b+\cdots.
\ee
It follows from (\ref{F}) and (\ref{SWA}) that
\be \label{SWF}
\Fh_{ab}\simeq F_{ab}-(F\th F)_{ab}-A_c\th^{cd}\nabla_d F_{ab}.
\ee

For an arbitrary scalar field $f(x)$
which is invariant under the gauge transformations,
we can make it covariant by modifying it to be
\be
\hat{f}\simeq f-A_a\th^{ab}\del_b f
\ee
so that
\be
\hat{\d}\hat{f}\simeq (\hat{f}, \hat{\lam}).
\ee
In general,
for an arbitrary differential form $\om$
invariant under gauge transformations,
\be
\hat{\om}_{abc\cdots}\simeq\om_{abc\cdots}
-A_d\th^{de}\nabla_e\om_{abc\cdots}
\ee
is covariant, i.e.
\be
\hat{\d}\hat{\om}\simeq(\hat{\om}, \hat{\lam}).
\ee

The Seiberg-Witten map is a crucial part in showing
the background independence of string theory.
In the next section we will define the NCYM action
and in sec.\ref{DBI} we will use the Seiberg-Witten map
to show that the DBI action with $B$ field background
is approximately equal to the NCYM action
on the noncommutative space with $\th\simeq B^{-1}$.

\subsection{NCYM}

In this section we construct
the noncommutative Yang-Mills (NCYM) action.

Denote the metric on the noncommutative space by $G$.
If $G$ is not constant, we have to modify it to
be a covariant metric for the NCYM action.
Let
\be
\nabla_a G^{bc}\equiv \del_a G^{bc}+G^{bd}\G^c_{ad}+\G^b_{ad}G^{dc},
\ee
then
\be \label{SWG}
\hat{G}^{ab}\equiv G^{ab}-A_c\th^{cd}\nabla_d G^{ab}+\cdots
\ee
is covariant in the sense that it transforms like $\hat{F}$
\be
\hat{\d}\hat{G}^{ab}\simeq
\th^{cd}\nabla_c\hat{G}^{ab}\del_d\hat{\lam}.
\ee

Define the NCYM action by
\be \label{NCYM}
S_{NCYM}=-\frac{1}{4g_{YM}^2}
\la tr(\hG\ast\Fh\ast\hG\ast\Fh)\ra
=-\frac{1}{4g_{YM}^2}
\la\hG^{ab}\ast\Fh_{bc}\ast\hG^{cd}\ast\Fh_{da}\ra.
\ee
To the lowest order in $\th$,
\be \label{action}
\hat{\d}S_{NCYM}
=-\frac{1}{4g_{YM}^2}\la(tr(\hG\Fh\hG\Fh),\hat{\lam})\ra=0.
\ee
So the NCYM action is invarinat under gauge transformations.

The NCYM action can be further simplified at the Poisson level.
To do so we derive an identity that will also be used later.
First consider the integration
\be
\la tr((M,N)MN)\ra,
\ee
where $M, N$ are symmetric or antisymmetric matrices of functions.
Using (\ref{cyclic}), we find
\bea \label{MNMN}
\la tr((M,N)MN)\ra &=& \left\la tr\left( (M,NMN)
-\th^{ab}\del_a MN\del_b MN
-\th^{ab}\del_a MNM\del_b N\right) \right\ra \nn\\
&=& -\left\la tr\left(\th^{ab}(\del_b NMN\del_a M)^T\right)\right\ra \nn\\
&=& -\la tr((M,N)MN)\ra = 0.
\eea
It follows from this that
\bea
\la tr(M\ast N\ast M\ast N)\ra
&\simeq& \left\la tr\left(
(MN+\frac{i}{2}(M,N))\ast(MN+\frac{i}{2}(M,N))\right)\right\ra \nn\\
&=& \left\la tr\left( MNMN+\frac{i}{2}(MN,MN)+i(M,N)MN\right)\right\ra \nn\\
&=& \la tr(MNMN)\ra.
\eea
(In fact, one can show that
\be
\la tr(M\ast N\ast M\ast N')\ra \simeq \la tr(MNMN')\ra.
\ee
for symmetric or antisymmetric matrices $M,N,N'$.)
Hence, the NCYM action at the Poisson level is just
\be
S_{NCYM}\simeq -\frac{1}{4g_{YM}^2}\la tr(\Gh\Fh\Gh\Fh)\ra.
\ee

\subsection{NCYM and DBI} \label{DBI}

In the Seiberg-Witten limit,
it is generally believed that the low energy effective field theory
for a D-brane in the background of a nonconstant $B$ field
lives on a noncommutative space
with the Poisson structure $1/B$ \cite{CF,HY}.
It is therefore natural to propose that the NCYM action
describes the physics for the $U(1)$ gauge field on the D-brane.
On the other hand, the DBI action is known to be the low energy
effective action for slowly varying fields on a D-brane \cite{FT,Leigh}.
We should thus be able to match the NCYM action with the DBI action
in the leading order in $\a'$ at least at the Poisson level.
This was shown for constant $B$ in \cite{SW,Cor},
and for generic $B$ in \cite{JSW} in the language of
covariant coordinates.
Here we want to rederive this result for nonconstant $B$
in the language of covariant derivatives.

Derivative corrections to the DBI action \cite{Tseyt,AT,And}
is of order ${\cal O}(F^2(\del F)^2)$ for bosonic string theory
and of order ${\cal O}(F^2(\del^2 F)^2)$ for superstring theory.
These are subleading terms in $\a'$ and thus should be ignored.
Since $F$ can only appear via the gauge invariant
combination $(B+F)$ in the D-brane action,
this means that terms linear in $(\del B)$
should cancel in the NCYM action.

The DBI action of a D-brane is
\be
S_{DBI}=T_p\int d^D x \sqrt{g+2\pi\a'(B+F)},
\ee
where $g$ is the closed string spacetime metric,
and $B$ is the NS-NS $B$ field background.
The D$p$-brane tension is
\be
T_p=\frac{1}{(2\pi)^p g_s(\a')^{(p+1)/2}}.
\ee
We will identify the $F$ in the DBI action with
the commutative $F$ related to $\hat{F}$
in the Seiberg-Witten map (\ref{SWA}).

Using the identity
\bea
\sqrt{det(1+M)}&=&1+\frac{1}{2}tr(M)
+\frac{1}{8}(tr(M))^2-\frac{1}{4}tr(M^2) \nn\\
&&+\frac{1}{48}(tr(M))^3-\frac{1}{8}tr(M)tr(M^2)+\frac{1}{6}tr(M^3)
+{\cal O}(M^4),
\eea
one finds that the leading order terms in $\a'$
in the DBI Lagrangian are \cite{SW}
\be \label{DBIL}
{\cal L}_{DBI}=(2\pi\a')^{(p+1)/2}T_p|Pf(B+F)|-
\frac{1}{4}(2\pi\a')^{(p+5)/2}T_p|Pf(B)|(f_1+f_2),
\ee
where $G$ and $\th$ are the symmetric and antisymmetric part
of $(g+B)^{-1}$, i.e.,
\be
\frac{1}{g+2\pi\a'B}=G+\frac{1}{2\pi\a'}\th,
\ee
and
\bea
f_1 &=& (1+\frac{1}{2}tr(\th F))tr(GFGF)-2tr(\th FGFGF)+\cdots, \\
f_2 &=& (1+\frac{1}{2}tr(\th F))tr(GBGB)-2tr(GBGF)+\cdots.
\eea
In $f_1$ and $f_2$ the omitted terms
are of higher order in $\th F$ or $\eps$
in the Seiberg-Witten limit (\ref{zsl}).
Since $B^{-1}\sim\th$, some terms omitted in $f_2$ are in fact
of the same order as some of the terms in $f_1$ which we keep.
However, as we have only defined the calculus at the Poisson level,
we can only compare terms of the leading order in $\th$,
and the background field $B$ is viewed as of order
${\cal O}(\th^0)$ in the sense of Poisson brackets.
Our result in this section will be justified in sec.\ref{mm}
by the background independence of the matrix model.

The first term in (\ref{DBIL}) is a total derivative
for $dB=0$ and $B>F$ in the sense that $|Pf(B+F)|=Pf(B+F)$.
The terms in $f_2$ are independent of $F$
up to a total derivative if $B$ and $g$ are constant,
but we need to keep it for nonconstant $B$.

Compared with the constant $\th$ case,
the NCYM for nonconstant $\th$ is roughly speaking
only modified by replacing all derivatives by
the covariant derivatives $\nabla$ as shown in
(\ref{SWF}), (\ref{SWG}).
Using the Leibniz rule for the covariant derivative, e.g.
\be
\del_a(f_b g^b)=(\nabla_a f_b)g^b+f_b(\nabla_a g^b),
\ee
we find after straightforward calculations that
$f_1$ agrees with the NCYM action for the coupling constant
\be
g_{YM}^2=(2\pi)^{(p-5)/2}g_s(\a')^{-2}.
\ee
In matching the two actions,
there is actually an ambiguity in choosing
whether the metric $G$ in (\ref{action})
is defined by $-B^{-1}\ast g\ast B^{-1}$ or just $-B^{-1}gB^{-1}$.
In our approximation, however,
there is no difference between these two choices.

To take care of $f_2$,
we need to modify the NCYM action as
\be \label{NCYM2}
S_D=-\frac{1}{g^2_{YM}}\la(tr(\Gh\ast(\Fh+\Bh)\ast\Gh\ast(\Fh+\Bh))\ra,
\ee
where
\be
\Bh_{ab}\equiv B_{ab}-A_c\th^{cd}\nabla_d B_{ab}+\cdots.
\ee
Since
\be
\hat{\delta}\hat{B}=(\hat{B}, \hat{\lam}),
\ee
$S_D$ is gauge invariant.
We will see later that this modification is needed for
the NCYM action to have a matrix model interpretation.

Let us call the metric appearing
in the fundamental string action the closed string metric,
and the metric appearing in the NCYM action
the open string metric \cite{SW}.
It is amusing to see that for the closed string,
$g_{ab}$ is the metric for the basis $dx^a$,
and $G^{ab}=-(B^{-1}gB^{-1})^{ab}$ is the metric for the basis $e_a$
\be
ds^2_{closed}=g_{ab}dx^a dx^b=G^{ab}e_a e_b.
\ee
On the other hand, it is the opposite for the open string.
$G_{ab}=-(Bg^{-1}B)_{ab}$ is the metric for the basis $dx^a$,
and $g^{ab}$ is the metric for the basis $e_a$
\be
ds^2_{open}=g^{ab}e_a e_b=G_{ab}dx^a dx^b.
\ee

\subsection{Matrix Model} \label{mm}

Due to (\ref{xi}),
we expect that a matrix model expression for
the NCYM is available for the basis of one-forms $e_a$.
Let
\be
\hat{A}=e_a\hat{A}^a, \quad
\hat{F}=\frac{1}{2}e_a e_b\hat{F}^{ab}.
\ee
It can be checked that
\be
\hat{F}^{ab}=(X^a, X^b)-P^{ab}(X),
\ee
where
\be
X^a\equiv \Phi^a-\hat{A}^a.
\ee
This $X$ is thus the so-called covariant coordinate \cite{Schupp}.

In terms of $X$
the action of the NCYM (\ref{NCYM2}) can be written as
\be \label{matrix}
S_{NCYM}=\frac{1}{4g^2_{YM}}
\la\hat{g}_{ab}\ast[X^b,X^c]_{\ast}\ast
\hat{g}_{cd}\ast[X^d,X^a]_{\ast}\ra.
\ee
This is precisely the leading order term one expects
for a matrix model in a curved space time \cite{Doug}.
The background independence of $B$ for the NCYM action
is manifest in this form,
as in the case of constant $B$ \cite{SW}.
In particular, for the case of flat spacetime,
we know that this expression is exact.
One can start from this action to obtain
the higher order terms in the NCYM action (\ref{NCYM2}),
using the Kontsevich formula to define the $\ast$-product.

\section{Remarks} \label{Remarks}

\subsection{Dual Description} \label{Dual}

Instead of using differential forms
to describe differential calculus,
we can just use functions and derivatives.
These two descriptions are dual to each other.
Starting with the Poisson differential calculus
in the previous section,
here we construct the noncommutative algebra
for derivatives and functions in the way
that has been used for many noncommutative spaces in the past
\cite{integral}.

Classically, when the exterior derivative $d$ acts on functions,
it is equivalent to $dx^a\del_a$.
Let us assume that this holds also for noncommutative spaces.
For example, we have
\be \label{dx}
dx^a=[dx^b\del_b, x^a]_{\ast}.
\ee
At the Poisson level, it is
\be
dx^a=-(dx^b p_b, x^a).
\ee
We will identify $p$ with the conjugate momentum of $x$.
It follows from this and (\ref{xdx}) that
\be \label{xp}
(x^a, p_b)=\d^a_b+\th^{ac}\G^d_{cb}p_d.
\ee

Assume that the Leibniz rule (\ref{Leib}) is observed,
for an arbitrary function $f(x)$,
the nilpotency of $d$ ($dd=0$)
will be satisfied at the Poisson level if
\be \label{pp}
(p_a, p_b)=-\frac{1}{2}r_{ab}^{cd}p_c p_d,
\ee
where $r$ is defined in (\ref{r}).

In the next subsection we will show that
both relations (\ref{xp}) and (\ref{pp})
are results of quantization for an open string
in the low energy limit.

\subsection{Open String Quantization} \label{open}

The bosonic action for an open string
ending on a D$p$-brane is given by
\be \label{string}
S=\int d^2\sigma\left(
g_{\mu\nu}\del_{\a}X^{\mu}\del^{\a}X^{\nu}
+2\pi\a'\eps^{\a\b}B_{ab}\del_{\a}X^a\del_{\b}X^b
\right),
\ee
where $\mu, \nu=0,1,\cdots,9$, and $a, b=0,1,\cdots,p$,
and $g$ is the closed string metric.
In the low energy limit,
the string is very short and so
one can approximate the string by a straight line
\be
X(\s)=x+\s X',
\ee
where $X'\simeq (X(\pi)-X(0))/\pi$ is very small.
For $H=0$, the symplectic two-form is \cite{HY}
\be
\Om=\hat{\cal F}_{ab}(X)dX^a dX^b|_{\s=0}^{\s=\pi},
\ee
where $\hat{\cal F}=B$ in the Seiberg-Witten limit.
>From this we can find the Poisson brackets among $x$ and $X'$,
and we will keep terms of order ${\cal O}((X')^2)$ or lower,
since we assumed that $X'$ is small.
This is equivalent to say that $B$ is slowly varying
compared with the length of the string,
and so we will only keep terms up to ${\cal O}(\del^2 B)$.
One then obtains
\bea
(x^a, x^b)&=&\th^{ab}(x) \nn\\
(y^a, y^b)&=&\th^{ab}(y)\simeq
\th^{ab}(x)+(y-x)^c\del_c\th^{ab}
+\frac{1}{2}(y-x)^c (y-x)^d\del_c\del_d\th^{ab},
%(x^{a},X'^{b})&=&-\frac{1}{\pi}B^{ab}\nn\\
%(X'^{a},X'^{b})&=&+\frac{1}{\pi}B^{ac}C_{cd}[B+\pi C]^{db}
\eea
%where $C=X'\partial B+(\pi/2)(X')^{2}\partial^{2}B$.
where $x\equiv X(\s=0)$ and $y\equiv X(\s=\pi)$.

The momentum associated with the endpoint at $\s=0$ is approximately
\be
p_a\simeq \pi\th^{-1}_{ab}(x){X'}^b.
\ee
Remarkably, within our approximation
we can calculate precisely the desired commutation relations
(\ref{xp}) and (\ref{pp}) among $x$ and $p$
for the connection one-form
\be \label{Gamma1}
\G^c_{ab}=\th^{cd}\del_a\th^{-1}_{db}.
\ee
Comparing this with (\ref{Gamma}),
we find that $x$ here is already
the canonical coordinate denoted $\Phi$ in sec.\ref{PDC}.
Since we ignored terms of order $\del^3 B$,
$\th$ is well defined only up to quadratic terms in $x$
as our $P$ in (\ref{phiphi}).
%Hence we calculate easily the commutator and anti-commutator 
%among $x$ and $p$ via (\ref{xp}), (\ref{pp}) and recover
%(\ref{xdx}), (\ref{r}) at the Poisson level.

It follows from (\ref{Gamma1}) that
the Leibniz rule (\ref{Leib}) is satisfied.
It is interesting to note that a priori
the Leibniz rule does not have to be satisfied
from the viewpoint of open string quantization.
(In the defining conditions for the Poisson differential calculus,
only the Jacobi identities for functions are guaranteed by
the associativity of a consistent quantization.
In fact, the Jacobi identities among differential forms are
not automatically satisfied either,
unless the exterior derivative is an element of
the algebra obtained from open string quantization.)
Nevertheless the result of open string quantization
is just sufficient to define the noncommutative gauge theory
and the Seiberg-Witten map.
The differential calculus defined in \cite{CH}
happen to be a natural framework for describing
a D-brane worldvolume in $B$ field background.

The matrix model action (\ref{matrix})
is greatly simplified when the metric $g$
is constant (and so the space is flat)
because $\hat{g}=g\;(\hat{g}=g(X)=g(\Phi -A))$ for this case.
For the connection (\ref{Gamma1}),
this implies that $\nabla G=0$ and thus $\hat{G}=G$.
Hence it is also the case when the NCYM action (\ref{NCYM})
is greatly simplified.

\subsection{Conclusion}

The property of background independence in $B$
is shown to hold perfectly for the NCYM action
in \cite{SW} for constant $B$.
This is because the NCYM action for constant $\th$
is equivalent to the matrix model action (\ref{matrix}) \cite{SW}.
Obviously that action is not changed under
\be
[x^a,x^a]=i\th^{ab}\rightarrow [{x'}^a,{x'}^b]=i\th'
\ee
as long as $x+\Ah=x'+\Ah'$.
(See also \cite{Seiberg} for relevant discussions.)
Since the NCYM action (\ref{NCYM2}) can be rewritten as
the matrix model action (\ref{matrix}),
obviously it is independent of the choice of $\Gamma$,
as long as (\ref{covar-th}) is satisfied.
(On the other hand, for the NCYM action (\ref{NCYM})
which does not have the background term,
it can depend on the choice of $\Gamma$.)
Note that this independence of $\Gamma$ is actually
associated with the symmetry of general coordinate
transformations on the brane.
If one quantizes the open string
in two different coordinate systems
and get two $\th$'s and two $\G$'s,
although the $\th$'s will be related by
a coordinate transformation,
the $\G$'s will not.
The fact that the D-brane theory should be
invariant under coordinate transformations on the brane
implies that the NCYM theory for D-brane should be
independent of the choice of $\G$.
This does not mean that we do not need $\G$---
at least we need it to satisfy (\ref{covar-th}).
Moreover, to deal with specific problems,
some choices of $\G$ may be more convenient than others.

The NCYM action for D-branes with nonconstant background
(\ref{NCYM2}) may be useful even for the flat case.
Note that for the quantization of the open string,
only the gauge invariant quantity $B+F$ matters,
although we have conventionally called the background part
of $B+F$ the $B$ field background.
Nevertheless if $F$ has a background value due to
some source living on the D-brane,
it will also make the D-brane worldvolume noncommutative.
For configurations with a background field which
varies greatly over the D-brane,
such as the configurations corresponding to
a semi-infinite string ending on the D-brane \cite{CM},
we expect that our description may be more appropriate
than the NCYM with constant $\th$,
since a small fluctuation of that configuration
is still a big fluctuation for any choice of constant background.
(On the other hand, as we just mentioned,
the action (\ref{matrix}) is background independent,
so nonperturbatively they must be equivalent.)

In this paper we constructed the noncommutative calculus
suitable for the description of D-branes,
and checked that the NCYM action matches with the DBI action
at the Poisson level for $H=0$.
It is natural to ask how these results
can be generalized when $H\neq 0$.
The (associative) noncommutative algebra
for this case was found in \cite{HY}.
The novelty of this algebra is that
the commutation relations between functions of $x$
is a function of both $x$ and $p$.
This poses a problem to the definition of
gauge transformations on such noncommutative spaces,
but this problem was solved in \cite{Ho2}.
Hopefully a differential calculus can be constructed so that
certain geometric understanding
about these gauge symmetries can be obtained.

\section*{Acknowledgment}

The author thanks Chong-Sun Chu, Miao Li and Jimmy Zung
for valuable discussions.
This work is supported in part by
the National Science Council,
the Center for Theoretical Physics
at National Taiwan University,
the CosPA project of the Ministry of Education,
and the National Center for Theoretical Sciences, Taiwan.

%\end{references}
%\end{narrowtext}
\end{document}